\documentclass[twocolumn,showpacs,preprintnumbers,amsmath,amssymb]{revtex4}

\usepackage{graphicx}
\usepackage{dcolumn}
\usepackage{bm}

\newcommand{\be}{\begin{equation}} 
\newcommand{\en}{\end{equation}}
\newcommand{\bea}{\begin{eqnarray}}
\newcommand{\ena}{\end{eqnarray}}

\newcommand{\hbo}{\hbox to 1 true cm {\hfill } } 
\newcommand{\tr}{\hbox{tr}}

\begin{document}

\preprint{KA-TP-04-2003 \hspace{1cm} UNITUE-THEP/7-2003}

\title{ Vortex critical behavior at the de-confinement  phase transition} 

\author{Kurt Langfeld}
 \email{kurt.langfeld@uni-tuebingen.de} 

\affiliation{%
Insitut f\"ur Theoretische Physik, Universit\"at Karlsruhe \\ 
D-76128 Karlsruhe, Germany. \\ 
and \\ 
Insitut f\"ur Theoretische Physik, Universit\"at T\"ubingen \\ 
Auf der Morgenstelle 14, D-72076 T\"ubingen, Germany. \\ 
}%

\date{\today}

\begin{abstract}
The de-confinement phase transition in SU(2) Yang-Mills theory 
is revisited in the vortex picture. Defining the world sheets of the 
confining vortices by maximal center projection, the percolation 
properties of the vortex lines in the hypercube consisting of 
the time axis and two spatial axis are studied. Using the 
percolation cumulant, the temperature for the percolation transition 
is seen to be in good agreement with the critical temperature of the 
thermal transition. The finite size scaling function for the cumulant is 
obtained. The critical index of the finite size scaling function is 
consistent with the index of the 3D Ising model. 
\end{abstract}

\pacs{ 11.15.Ha, 12.38.Aw, 12.38.Gc }
\keywords{Yang-Mills theory, confinement, center-vortices,
                             percolation,  finite size
                             scaling }
\maketitle

Understanding de-confinement at finite temperatures is one of the 
major challenges of QCD, the theory of strong interactions. 
Large scale collision experiments (RHIC) are currently operating 
in order to detect new states of matter at high 
temperatures~\cite{Heinz:2002gs,Kharzeev:2002hw}.  
The general believe is that the confinement of quarks is induced 
by the gluonic state, and that the mechanism can be anticipated 
within the pure $SU(N)$ gauge theory. Using dimensional reduction
in combination with lattice simulations of the emerging effective 
3D theory, it was recently shown~\cite{Kajantie:2000iz} 
that the non-perturbative effects of $SU(N)$ gauge theory 
are strong for temperatures which are accessible in collider
experiments. The universality class of the 
transition is of particular importance because the corresponding 
degrees of freedom are weakly interacting close to the transition, 
thus dictating the dynamics of the theory. 

\vskip 0.3cm 
Any theory which purports certain degrees of freedom 
of being relevant for confinement should also be able to explain 
de-confinement at finite temperatures. An example for such degrees of 
freedom are chromo-magnetic monopoles (a pedagogical review to
the subject can be found in~\cite{Chernodub:1997ay}). It was 
empirically observed that monopole currents, residing 
in the spatial hypercube, form a network which ceases percolation in the 
de-confined phase~\cite{Bornyakov:1991se,Damm:1997qx}. However, it 
seems likely that the monopoles are strongly coupled even close to the 
de-confinement transition. Indeed, arguments have been put forward 
that a (weakly interacting) 
Coulomb gas of chromo-magnetic monopoles is inconsistent 
with the phenomenology of confinement~\cite{Greensite:2003bk}. 

\vskip 0.3cm 
\begin{figure*}
\includegraphics[height=7cm]{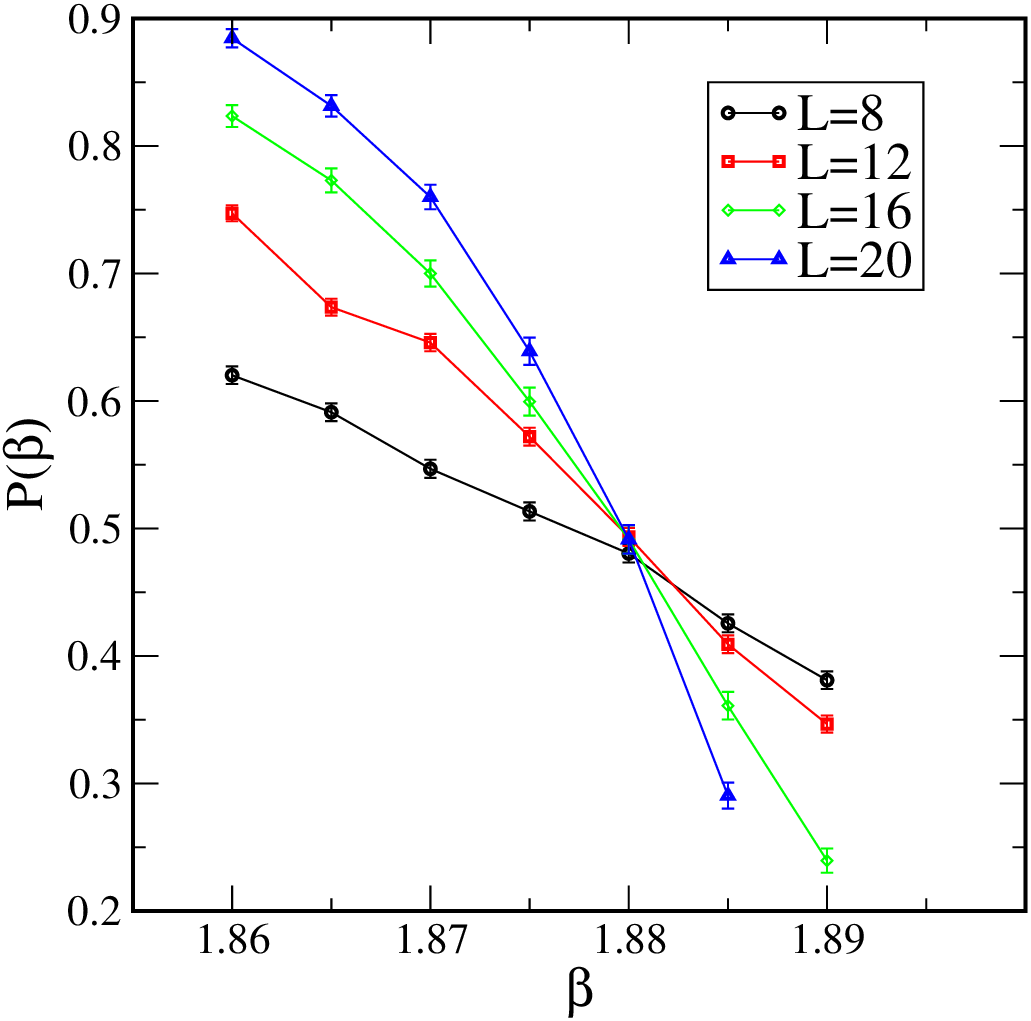} 
\includegraphics[height=7cm]{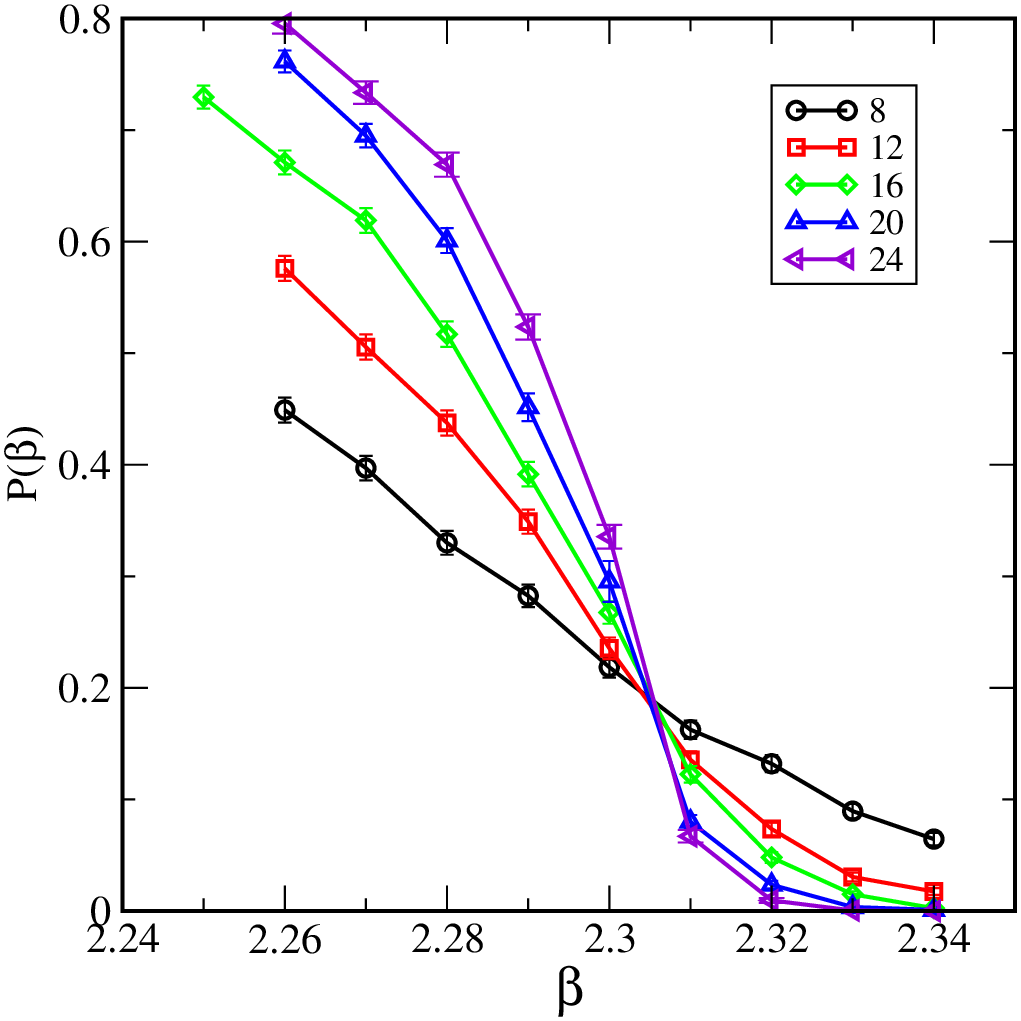}
\caption{\label{fig:1} The percolation cumulant $P(\beta )$ 
for several lattice sizes $L$ for $N_t=2$ (left) and $N_t=4$ (right).
}
\end{figure*}
Below, I will address two scenarios of the de-confinement phase
transition at finite temperatures. One of these exploits the realization 
of the global $Z(N)$ center symmetry of the corresponding 
$SU(N)$ Yang-Mills theories. 
The spontaneous breaking of this symmetry at high temperatures is 
signaled by a non-vanishing expectation value of the 
Polyakov line. In the case of the 
SU(2) gauge theory, the projection of the Polyakov line to the 
center elements has revealed that the effective theory of 
the corresponding 3D Ising model is short ranged at the 
transition~\cite{Fortunato:2000vf,Fortunato:2000ge,Satz:2001zf}.  
The thermal transition corresponds to a percolation transition 
of the site-bond clusters of the emerging 3D Ising model. 
More recently, it was observed~\cite{Engels:1998nv} that also the 
ratio of the Polyakov line critical amplitudes are in agreement 
with that of the 3D Ising model. 
These findings confirm that the transition of the SU(2) gauge theory 
belongs to the universality class of the $Z_2$ Ising model in 
three dimensions. 

\vskip 0.3cm 
A second scenario emerged with 
the advent of the so-called maximum center gauge 
(MCG)~\cite{Greensite:2003bk,DelDebbio:1996mh,DelDebbio:1997ke,DelDebbio:1998uu}: 
a tight relation between quark confinement and the vortex 
structure of the pure gluonic vacuum was established at a quantitative
level for $SU(2)$ and 
$SU(3)$~\cite{Faber:1999sq,Stack:sy} gauge theory: projecting 
$SU(2)$ lattice configurations onto those of a $Z_2$ gauge theory, 
the confining capabilities of the static quark anti-quark potential 
at large distances is retained. On the other hand, removing
the vortex structure from the lattice configurations ``by hand'', the 
static potential looses its linear rise and resembles a Coulomb 
potential. The key observation is that the vortices which emerge
from the above effective $Z_2$ gauge theory~\cite{DelDebbio:1998uu} 
are sensible degrees of freedom of continuum Yang-Mills theory: 
extrapolating to the continuum limit of vanishing lattice 
spacing, the planar vortex
density\cite{Langfeld:1997jx,DelDebbio:1998uu} as well as the binary 
vortex interactions~\cite{Engelhardt:1998wu} are finite in units of 
the fundamental energy scale (e.g.~string tension). These observations 
indicate that the MCG vortices are the degrees of freedom relevant 
for quark confinement. Recently, it was shown that the vortices 
might also play a role for other low energy observables: 
the removal of the vortex structure leads to a restoration of 
chiral symmetry~\cite{deForcrand:1999ms} and to a strong reduction 
of the infrared strength of certain Yang-Mills 
Greenfunctions~\cite{Langfeld:2001cz,Langfeld:2002dd}. 

\vskip 0.3cm 
The question whether one can grasp the essence of the de-confinement 
phase transition at finite temperatures within the vortex picture 
was firstly investigated in~\cite{Langfeld:1998cz,Engelhardt:1999fd}. 
One finds that the effective $Z_2$ theory which arises from the 
$SU(2)$ lattice gauge configurations after center 
projection~\cite{DelDebbio:1998uu} reproduces the right order 
of magnitude of the de-confinement temperature $T_c$. In order to 
detect the change of the vortex structure when $T_C$ is approached, 
the vortex world lines of a hypercube 
consisting of the time and two spatial directions were studied. 
While the vortices percolate at small temperatures, the vortices 
align parallel to the time axis direction for $T>T_c$. The 
de-confinement phase transition appears as vortex de-percolation 
transition. 

\vskip .3cm
A comparison of both scenarios of the de-confinement phase transition 
which have been discussed above suggests a certain duality: 
(i) the Polyakov line gives rise to a Ising model defined in the 
spatial hypercube; site-bond clusters start percolating in the 
high temperature phase; (ii) closed vortex loops emerge from 
MCG projection in the hypercube spanned by the time- and two spatial 
directions; the vortex clusters cease to percolate in the
de-confinement phase. In order to establish a duality between 
both pictures, one must verify that the critical behavior 
of the vortex percolation transition corresponds to the 3D Ising 
universality class.

\vskip 0.3cm 
In this letter, a detailed analysis of the center vortex loops 
in the time-like hypercube is performed at finite temperatures. 
Using the percolation cumulant, a precise determination 
of the critical de-percolation temperature $T_p$ is obtained. 
I find that $T_p$ nicely agrees with the thermal transition temperature 
$T_c$. Finally, the 
finite size scaling function of the percolation cumulant 
will show that the finite size scaling is consistent with 
that of the 3D Ising model. 

\vskip 0.5cm 
\begin{figure*}
\includegraphics[height=7cm]{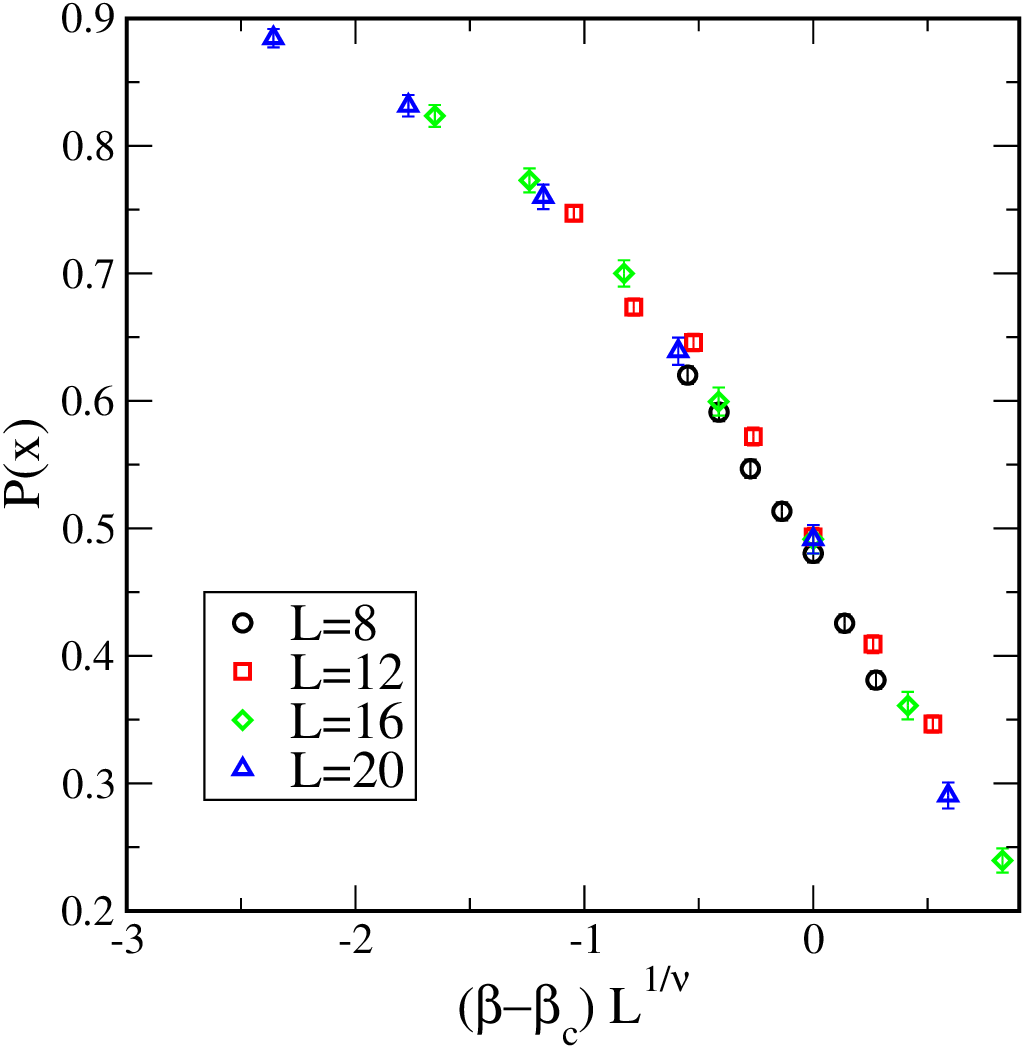} \hspace{0.5cm}
\includegraphics[height=7cm]{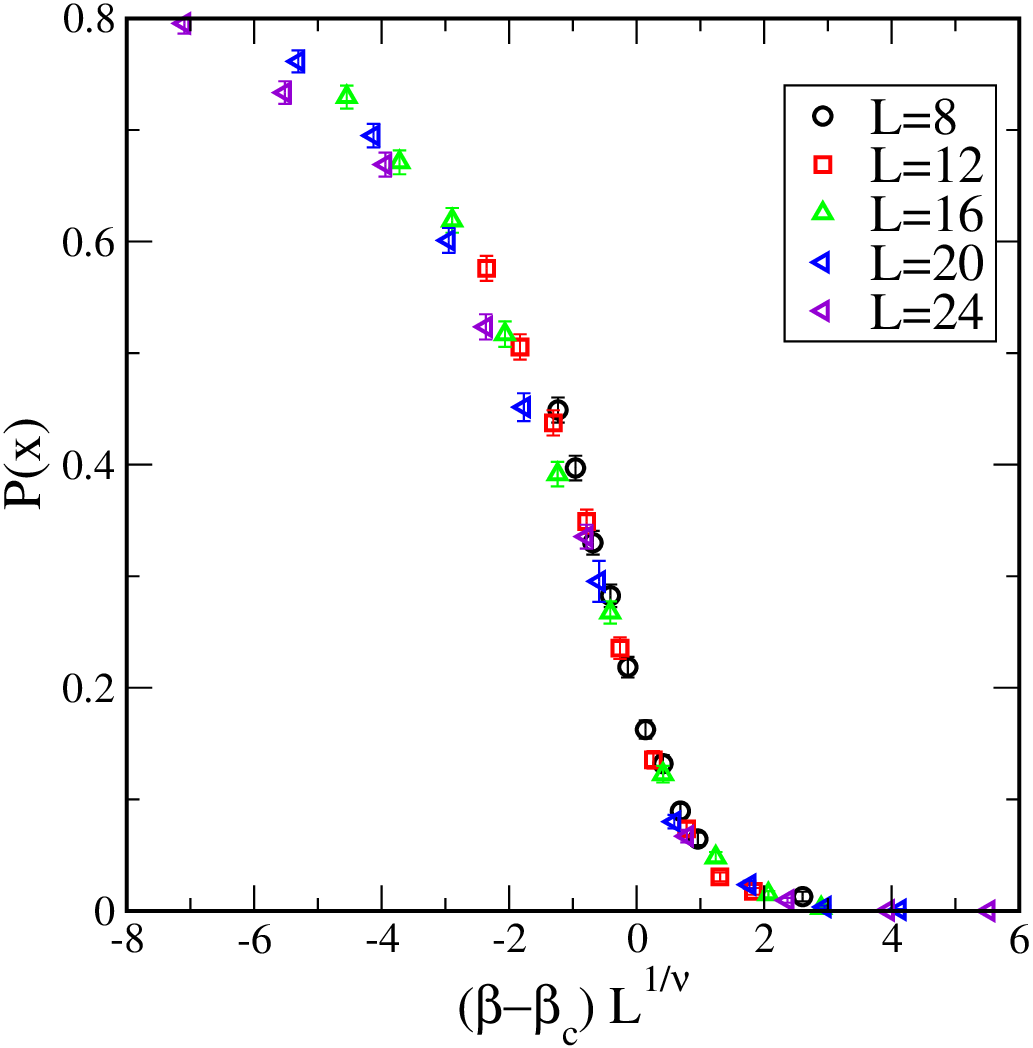}
\caption{\label{fig:2} The percolation cumulant $P(x )$ 
for several lattice sizes $L$ for $N_t=2$ (left) and $N_t=4$ (right) 
as function of the scaling variable $x$ with $\nu = 0.6294$ as 
input. }
\end{figure*}
The links $U_\mu (x)$ representing the $SU(2)$ gauge field
configurations are generated with the standard Wilson action. 
The Maximal Center Gauge is installed by maximizing the functional 
\be \label{eq:1} 
\sum _{x, \mu } \; \Bigl[ \tr \, U^\Omega 
_\mu (x) \Bigr]^2 \; \stackrel{ \Omega
}{ \rightarrow } \; \hbox{max}  \; , 
\en 
where $U^\Omega _\mu (X) = \Omega (x) \, U_\mu (x) \, \Omega ^\dagger 
(x+\mu) $ are the gauged links. Finding the global maximum of
(\ref{eq:1}) is a highly non-trivial task and is subject to Gribov 
ambiguities. One observes that bulk properties of the vortices 
are little affected by the choice of algorithm (see
e.g.~\cite{Langfeld:2001cz}). Here, I will not further address the 
Gribov problem, but will adopt a practical point of view: I will 
employ the procedure described
in~\cite{DelDebbio:1998uu} because it yields good scaling 
properties of the vortex matter. One must, however, keep in mind that 
it is not clear to which extent the choice of algorithm affects the 
results presented here. Center projection is then performed by 
\be \label{eq:2} 
SU(2) \rightarrow Z_2 : \; \;  U^\Omega _\mu (x) \rightarrow 
Z_\mu (x)  = \hbox{sign } \, \tr \, U^\Omega _\mu (x) \; . 
\en 
Defining 
\be \label{eq:3} 
v(p) \; := \; \prod _{l \in p } Z_l \; , \; \; l=\{x,\mu \} \; , 
\en 
one says that a vortex pierces an elementary plaquette $p$ if 
$v(p)=-1$. Hence, vortices exists on the dual lattice. 
In view of the $Z_2$ Biancchi identity, 
$$
\prod _{p \in c } v(p) \; = \;1 \; , 
$$ 
where $c$ denotes an elementary cube, 
vortices form closed world sheets in 4 dimensions and appear 
as closed loops in a particular 3D hypercube. In the following, we will 
consider the time-like hypercube spanned by the time- and two spatial 
directions, and we will investigate the properties of the closed 
vortex loops within this hypercube. A vortex cluster is defined as 
the vortex material which is connected by a single vortex world line. 
The time-like hypercube contains $L^2 \times N_t$ lattice points. 
The temperature $T= 1/N_t \, a(\beta) $ is adjusted by varying $\beta
$. Below, results for $N_t=2$ and $N_t=4$ are presented. 

\vskip 0.5cm 
For an investigation of the vortex percolation within the time-like
hypercube, one determines the minimal box which completely contains 
a given vortex cluster. If this box is as big as the time-like
hypercube, the particular cluster is said to percolate. In order to 
study the vortex critical behavior at the critical temperature, 
one defines the so-called {\it percolation
cumulant}~\cite{Fortunato:2000vf} $P(\beta )$, which is given by 
the average number of percolating vortex cluster divided by the number 
of configurations. Analogous to the Binder cumulant, the percolation 
cumulant is independent of the extent of the lattice $L$ at the critical 
point $\beta _c$. This implies that the curves $P(\beta )$
corresponding to different lattice sizes cross at the critical 
point. My numerical results are shown in figure \ref{fig:1}. 
I find 
$$ 
\beta_c = 1.880(3) \; \; \; (N_t=2), \; \; \; 
\beta_c = 2.305(3) \; \; \; (N_t=4). 
$$ 
These values are in reasonable agreement with those quoted for 
the  phase transition signaled by the percolation of the 
Polyakov line~\cite{Fortunato:2000vf}, i.e.,
\begin{eqnarray} 
\beta^{pol}_c &=& 1.8734(2) \; \; \; \; \; \; \, (N_t=2), 
\nonumber \\ 
\beta^{pol}_c &=& 2.29895(10) \; \; \; (N_t=4). 
\nonumber
\end{eqnarray}

As long as finite size scaling holds, the percolation cumulant 
should be a universal function of 
$$
x:=(\beta - \beta _c) \, L^{1/\nu } \; , 
$$
where $\nu $ is characteristic for the universality class of the 
phase transition. Figure \ref{fig:2} shows the vortex percolation 
cumulant as function the scaling variable $x$ for $N_t=2$ and 
$N_t=4$. The numerical results are consistent with the critical 
index $\nu = 0.6294$ of the the 3D Ising model. It is remarkable that 
the finite size scaling works for a wide range of the scaling variable
$x$. 

\vskip 0.3cm 
In conclusions, the picture of the de-confinement phase transition at 
finite temperatures as a vortex de-percolation transition is refined: 
the finite size scaling function of the vortex percolation cumulant 
is obtained. The critical index $\nu $ is seen to be consistent with 
the index of the 3D Ising model. Hence, the center vortex picture 
of the phase transition is put on equal footing as the percolation 
picture of site-bond clusters defined by the Polyakov 
line~\cite{Fortunato:2000vf}-\cite{Satz:2001zf}. Since the definition 
of the Polyakov line breaks covariance, the latter picture is
necessarily restricted to temperatures of the order of the de-confinement 
temperature. By contrast, the center vortex clusters are consistently
defined with the symmetries at all temperatures, and provide 
a picture dual to the Polyakov line picture at $T\approx T_c$. 
However, one must keep in mind that the Gribov problem renders 
an unambiguous definition of the vortex cluster cumbersome at the 
present stage of research. 

\par\bigskip
    
\noindent {\bf Acknowledgments:}
I thank M.~Quandt for helpful comments on the manuscript.

\end{document}